# Predictive drift compensation of multi-frame STEM via live scan modification

Matthew Mosse[1,2] *, Jonathan J.P. Peters[1,2], Eoin Moynihan[4], James A. Gott[5], Ana M. Sanchez[4], Michele Conroy[6], Lewys Jones[1,2]

[1]*Advanced Microscopy Laboratory, CRANN, Trinity College Dublin, The University of Dublin, Dublin, Ireland.*
[2]*School of Physics, Trinity College Dublin, The University of Dublin, Dublin, Ireland.*
[4]*Department of Physics, University of Warwick, Coventry, UK*
[5]*Advanced Materials Manufacturing Centre, Warwick Manufacturing Group, University of Warwick, Coventry, UK*
[6]*Department of Materials, Royal School of Mines, Imperial College London, Prince Consort Road, London, UK*

\* corresponding author: mossem@tcd.ie

## ABSTRACT

Scanning transmission electron microscopy (STEM) is widely used tool for materials characterisation. However, being a scanned technique, STEM is susceptible to sample, stage or beam drift, manifesting as distortions within images or movement in the field-of-view during multi-frame imaging. Often this is corrected post-acquisition using image registration of multiple frames, but drift reduces the usable area common to all frames. Here we present a method to mitigate sample drift by analysing past frames to predict the sampling-grid points for the immediately future frame. We present this correction across two time-scales and two length-scales. By offsetting the scan-grid framewise we remove long-range drift, and offsetting pixelwise we minimise intra-image warping. Examples are presented for both atomic-resolution imaging and lower-magnification in-situ video capture. The framework is general to raster, serpentine, interlaced and other scan patterns, as well as sequential or scan-rotation series STEM.

## INTRODUCTION

Scanning transmission electron microscopy (STEM) allows materials to be imaged across many length scales. STEM data is fundamentally a time series of intensity values which are placed into two-dimensional arrays to form images. If there are any displacements between the presumed position of either the beam or sample during the acquisition, distortions within the image may appear. Displacements may arise from effects such as thermal expansion of the column, sample or sample holder; in response to air pressure changes and flexing of the rubber seals; electrical noise, transient magnetic fields or fluctuating sample charging [1] [2]. Distortions may then appear as the image bending, scan line tearing, or corruption of spacings and angles.

One method to reduce the appearance of scanning artefacts is to acquire faster frames, averaging over multiple frames to regain the signal-to-noise ratio (SNR) [3], [4], [5]. For a given fixed total acquisition time, fractionating this time into multiple passes causes distortions (originating in the time domain) to be spread spatially across multiple frames, reducing the effect on any individual frame. Translational drift, or very low frequency distortions during that total time, now appear in the image series as shifts between frames and affine skews within each frame. Previously higher frequency intra-frame distortions (discontinuous tears/slices between scan-lines) now appear as continuous slow curvatures within frames which can be more easily corrected in post-processing [5], [6], [7]. Splitting the electron dose over many frames may confer additional benefits such as reducing beam-induced damage [8] as time is allowed for thermal/charge dissipation, diffusion of lost species, or bond relaxation [9]. Multi-frame STEM acquisitions are also indispensable for dynamic in-situ imaging of samples, to observe phase changes [10], radiation-induced effects [11] and the movement of atoms [12], defects [13], and domains [14] [15].

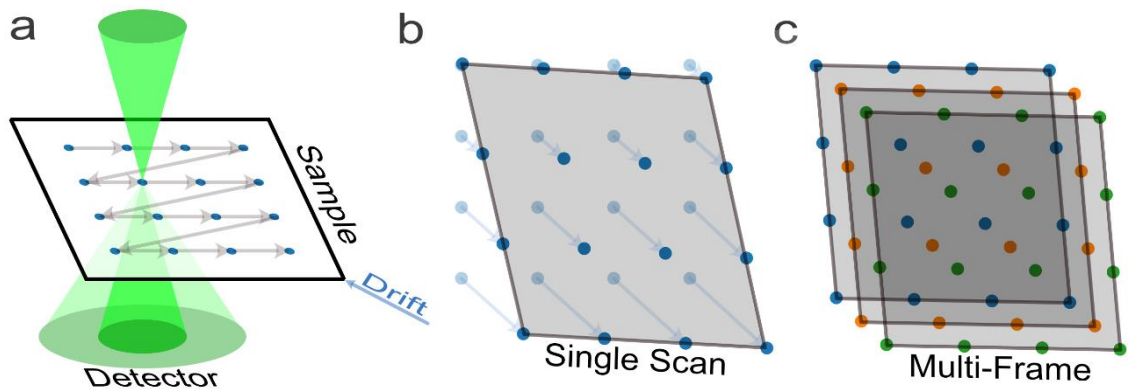

*Figure 1: (a) A raster scan from the perspective of the microscope scan-system, ignorant to any possible drift. (b) A single slow scan with the sample drifting up and to the left, will cause recorded image pixels to appear sheared to the bottom-right. Arrowheads indicate the increasing displacement with respect to time as the acquisition progresses. (c) This distortion can be reduced using multiple fast frames, but the usable common sample-area is reduced.*

Despite the reduced drift distortions in multi-frame imaging, translational drift especially will result in the region or feature of interest moving out of the field-of-view (FoV) as the acquisition proceeds. When a multi-frame stack of images is aligned in post-processing, the common FoV is reduced, as seen in Fig. 1, with the non-common areas typically cropped and discarded. Irradiation that does not yield useful data, wastes both time and is inherently dose inefficient [16]. Even irradiation outside the aligned common FoV may cause damage which can encroach into the final imaged area [9] [17].

Long acquisitions with large amounts of drift are common in light microscopy on biological samples; this can be solved by adjusting the position of the stage between frames to compensate for the drift [18]. In the case of electron microscopy, using physical stage movement is often not precise enough due backlash (mechanical hysteresis) and/or slip-stick movement (juddering) [19]. Using a piezoelectric stage may be sufficient in limited cases [20], but being limited in their travel range to approximately 1 μm before regular a mechanical stage is again needed. Live drift compensation using rigid offsets has been discussed in scanning electron microscopy (SEM) such as in recent work by Liu et al. [21]. However, their methodology is insufficiently described to reproduce, for example any details of drift fitting or prediction, or how they were able to predict the scan-offset for the second frame with only one preceding frame.

For STEM, the generally higher magnification than SEM, means even minor disturbances cause large image artefacts. Correcting the stage manually is occasionally necessary during long in-situ acquisitions, but this can be a crude adjustment. Perhaps the first to demonstrate drift-tracking (without prediction) was Kimoto et al. [22]. Later affine-skew analysis was identified by Sang and LeBeau as a future means to predict stage drift [4].

It is common practice for methods that require slow scans of minutes to hours, such as electron dispersive x-ray spectroscopy (EDX), electron energy loss spectroscopy (EELS), or four-dimensional STEM (4DSTEM), to use an additional signal and scan area to compensate sample drift. Typically this signal is the annular dark field (ADF) image acquired as fast scans on some other part of the sample [23]. This can be done every line or every few lines for a single slow scan, or every frame if many frames are being acquired and summed, as done in some commercial software [24] [25] [26]. Scanning extra purely for drift correction is dose and time inefficient. Our approach described here differs by not requiring any additional signal or scan area, i.e. it is scanning grid recursively correcting itself.

Here we demonstrate an open-source, reproducible, method to counter drift *before* it corrupts the acquisition, with no additional sub-scan area or unused frames. Drift-rate is continuously assessed and refined as new frames are acquired, adapting to evolving drift behaviours. Scan-grid offsets are pre-calculated for the upcoming frame yet to be recorded. The prediction and compensation can be performed on a whole-frame basis to compensate rigid translation preventing features of interest moving within the field-of-view. Further, we also introduce non-linear and pixelwise predictions for the first time, allowing compensation of time-varying drifts responsible for image warping.

## METHODS

To reliably identify and eliminate drift live during the acquisition, we need both an accurate registration method for captured frames, and a means of predicting the drift at a given future time. The conceptual architecture of our method is shown in Figure 2.

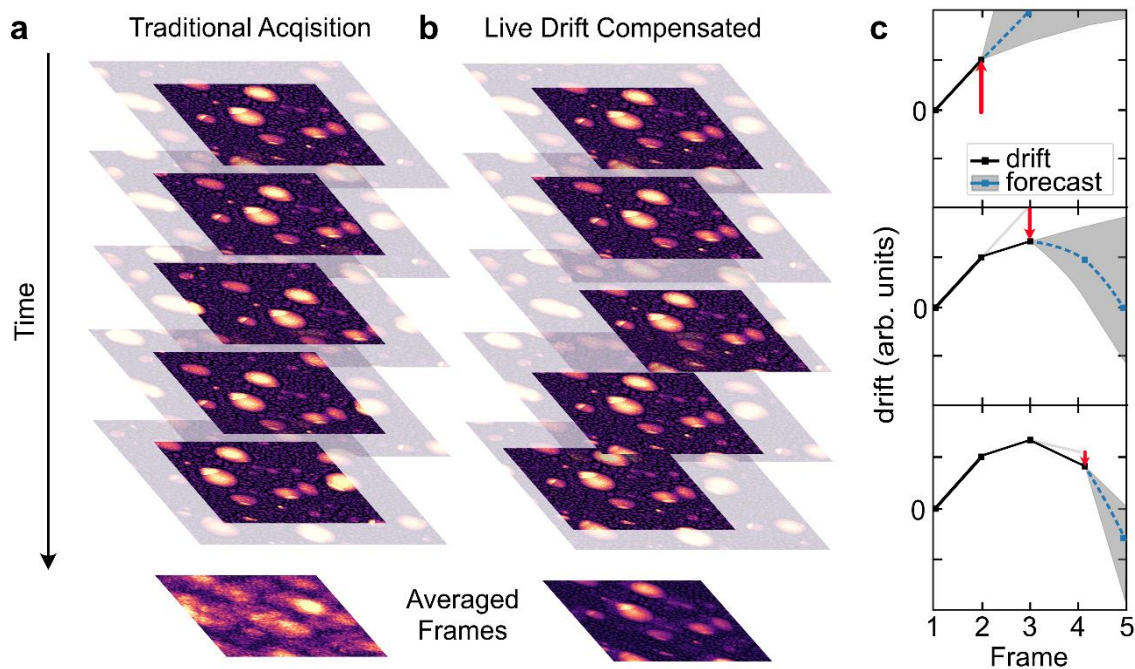

*Figure 2: (a) A series acquisition showing sample drift leading to a misaligned average frame with reduced common area once aligned. (b) Series acquisition with drift measured (from frame two onwards) and predictively compensated before each subsequent scan, preserving the maximum attainable FoV with no post-processing needed. (c) Schematic showing evolution of the drift prediction, dynamically responding to changes in drift rate/direction, as each subsequent frame contributes to the fit.*

## IMAGE REGISTRATION

There are many image registration methods that could be chosen to observe the drift rate. Here we chose cross-correlations as they are efficiently calculated in Fourier space, do not require manual labelling for feature tracking, and are relatively robust to low signal-to-noise ratios (SNR) [6]. No assumptions were made about the content of the images, and so no high-contrast markers (fiducials) were assumed to be available for registration, as is often the case in light microscopy [18], [27].

One issue that may arise in periodic images (e.g. single crystals), is the risk of an unwanted miscorrelation to an equivalent symmetry point, i.e. 'unit-cell hopping'. This can be improved by using slight variations of cross-correlation such as mutual-correlation [28] or phase-correlation [29]. This may still fail if the data is extremely noisy (which is common in fast frames) or highly uniform (as is common in crystalline samples). For fast multi-frame acquisitions, where only slight shifts are expected between consecutive frames, it is reasonable to constrain the correlations to the first unit-cell. This can be done in a computationally efficient way by applying a smooth-edged mask within the Fourier multiplication step. Additionally, when using this live drift compensation, the field of view is constantly recentred. This ensures total drift remains under one unit cell between the first and last frame, and further justifies the use of a unit-cell mask.

## PREDICTION

Any mathematical extrapolation using only two data-points, and without prior knowledge, can only predict linear trends. In our case, having observed only two frames, the scan offset of frame three (from frame one), is inferred by applying twice the measured offset of frame two from frame one. Once acquired, the third frame is then correlated and the current drift-rate is updated. As additional drift-rate observations accrue, a more and more accurate fit and extrapolation can be made, for example with a quadratic (or higher order) fit, rolling average, or other methods discussed later.

For low-dose noisy frames, it is advantageous to make multiple, independent, measures of displacement. For a frame-series of length N, we can correlate any pair of images which yields an N*N array. This array is a heavily over-defined matrix, where for efficiency, we need not calculate the diagonal (images with themselves), or the lower triangle (the negative of the pairwise correlations in the upper triangle). In the further interests of computation speed, frames far separated in time need not be compared. This is especially true in the case of in-situ measurements, where the nature of the sample may have changed significantly. This is discussed further in Supplementary Fig. S1.The remaining, relevant displacement, can be solved using a holistic least-squares solution [30], which provides a more noise resilient measurement of the drift behaviour across the series.

To predict the drift that will occur in the immediately next frame, some function must be fit to the past displacements. Since multiple factors can affect the duration of frame capture, e.g. computer/readout overhead, data buffering, variable pixel dwell-time, or variable scanning-path strategy, the interval between fames may not be uniform, and frame number alone may be insufficient. In our approach, we record the start-time of every frame $t_n$, defined absolutely relative to the control-PC's hardware clock.

For a prediction to be practical, we assume drift is not entirely random and can be fitted. For fast scans, and when comparing a reasonable number of recent frames, this snapshot of time can be approximated by a simple polynomial. For a robust extrapolation, the appropriate polynomial order is dependent on the number of unique time-points forming the prediction. Smooth fits to the drift are physically realistic since, stages/holders have some finite inertia, and in the case of thermal expansion, we expect a gradual and monotonic settling toward thermal equilibrium, which is often linear [7] or smoothly varying [4]. Where drift-rate evolves over time, the most recent observations are likely the most relevant. With this, the fit can optionally be weighted with some function diminishing into the past (e.g. exponential decay), which favours newer observations.

For slower frames, or acquisitions with more complex drift behaviour, more complex prediction functions can be employed. One example is autoregressive integrated moving average (ARIMA), which is a class of prediction models used for time-series data with periodic variations and trends [31], which describes low frequency fluctuations occurring during a STEM acquisition, and a modified seasonal ARIMA (SARIMA) can be used to model periodic changes.

Two separate functions in x and y, $f_x$ and $f_y$ respectively, are used as the drift in the two directions can be treated independently [7]. For purely translational drift, predictions can be made for the start time of the next frame $t_{n+1}$ and applied as a transformation of the *whole frame*, meaning that for every pixel in row $i$ and column $j$ has the same applied offset $dx_{i,j}$

$$dx_{i,j} = f_x(t_{n+1}) \qquad (1.1)$$

$$dy_{i,j} = f_y(t_{n+1}) \qquad (1.2)$$

Alternatively, the prediction can be implemented pixel-wise to correct image distortions arising within the frame-time due to non-linear drift. To achieve this, the predicted drift at each pixel's acquisition time can be calculated and applied to each pixel position, such that

$$dx_{i,j} = f_x(t_{n+1} + t_{i,j}) \tag{2.1}$$

$$dy_{i,j} = f_y(t_{n+1} + t_{i,j}) \tag{2.2}$$

where $t_{i,j}$ is the start time of each pixel relative to the start of it's parent frame, and is given by

$$t_{i,j} = \sum_0^i (T_{line}) + T_{LFB} + \sum_0^j (\delta_t) \tag{3}$$

In this expression, $T_{line}$ is the slow scan line time including the line flyback time $T_{LFB}$, and $\delta_{px}$ is the pixel dwell time[16]. By applying predictive drift compensation individually to the pixel positions (times), artefacts in the image such as affine shear or even nonlinear distortions can be removed before they effect the image.

In this present work, registrations were done on a frame-by-frame basis, and therefore correction of distortions were only predicted and compensated at or below this frequency. However, an implementation that takes intra-frame distortion into account may be possible if these higher frequency variations are predictable [2], though issues with synchronisation may occur and may be more computationally taxing.

IMPLEMENTATION

In our method, all frames are treated as part of the acquisition, but the second frame cannot be compensated since this frame is used for the initial prediction of drift velocity. However, an implementation that takes into account drift during a pre-acquisition navigation time is possible (see supplementary Fig. S2).

One requirement of using the scan-deflectors to track a moving specimen, is that they should not already be being driven at 100% of their maximum rated power. This is in fact already the common practice, as a pre-scan is often used during for frame fly-back [16], [32]. This occurs at the start of each line, meaning that the imaged area is already reduced relative to the total possible deflector strength. If the scan area is, say, 50% of the total possible deflection strength, our method could accommodate half a frame of drift during the live compensated acquisition and remain within the safe coil limits. Since the scan-coils are able to function from low magnifications and higher (large coil currents and beam deflections), no hardware limitations are expected at anything but the lowest magnifications an this caveat is not really a limitation.

A limitation in the case of ARIMA is that data and predictions must have equal time intervals. If there are different times between frames (as is the case here due to the computation) or between predictions (as is the case for pixel-wise prediction) a Kalman filter could be used on both input data and output predictions to produce predictions at arbitrary time steps [33]. For the sake of computational speed, a polynomial fit was used for any pixelwise compensation presented in this paper.

The methods explained here are general enough to be applied on any scan generator, such as Nion Superscan, JEOL MDP, Gatan DigiScan 3, and point electronic REVOLON. These are capable of a full pixel-by-pixel compensation, but others may be capable of framewise compensations. For the results shown here, we used a point electronic REVOLON which can deliver fully arbitrary scans.

# RESULTS AND DISCUSSION

*(1) Synthetic Acquisition Testing*

To evaluate the accuracy of our proposed method, it was compared with the current state of the art, which is a multi-frame acquisition followed by post-processing alignment in the program SmartAlign [5]. Frames were first captured, with no drift-prediction used, using a ThermoFisher Spectra Ultra 300, of rhombohedral barium titanium oxide (BTO) in the (pseudo cubic) ⟨100⟩ zone axis under cryogenic conditions. During such cooled (or heating) experiments it is typical to observe a non-negligible amount of drift due to thermal instability.

Frames were captured at 1024x1024 pixels, and a central 512x512 sub-region of the image is used which never leaves the experimental field-of-view during the recording. This experimental image stack can then be used to produce a realistic synthesised acquisition with a known grown truth drift profile as described below. If no compensation is used, Fig. 3a shows the loss of resolution due to blurring and 3e shows the loss in common area.

The proposed drift compensation algorithm was run as if the drift were occurring live by measuring historical drift, then predicting and compensating (offsetting) the location of the sub-region on the next frame (made possible since the experimental data spanning 1024 px provides padding for the 512 px synthetic acquisition). Fig. 3d shows the prediction's compensated displacement vectors and, for comparison, those measured by SmartAlign which are considered as the ground truth. Two different prediction functions were compared on this synthetic acquisition; ARIMA and polynomial fitting.

Fig. 3h shows the residual displacement relative to the ground-truth, and it was found that the ARIMA forecasting method produced the lowest average error of 2.4 pixels, compared with 3.4 pixels for the simple polynomial fit. An assumption of the ARIMA method is that the input data and output predictions have equal time steps, which was true for the experimental multi-frame acquisition in this test. However, in a genuine experimental implementation of our live drift compensation, the prediction calculation would be done between frames, meaning slightly varying times between frames cannot be ruled out. For this reason, hereafter we use a polynomial fit for predictions, as it was nearly similar in accuracy as ARIMA but did not have the additional complexity and restrictions should unequal frame-intervals occur in future.

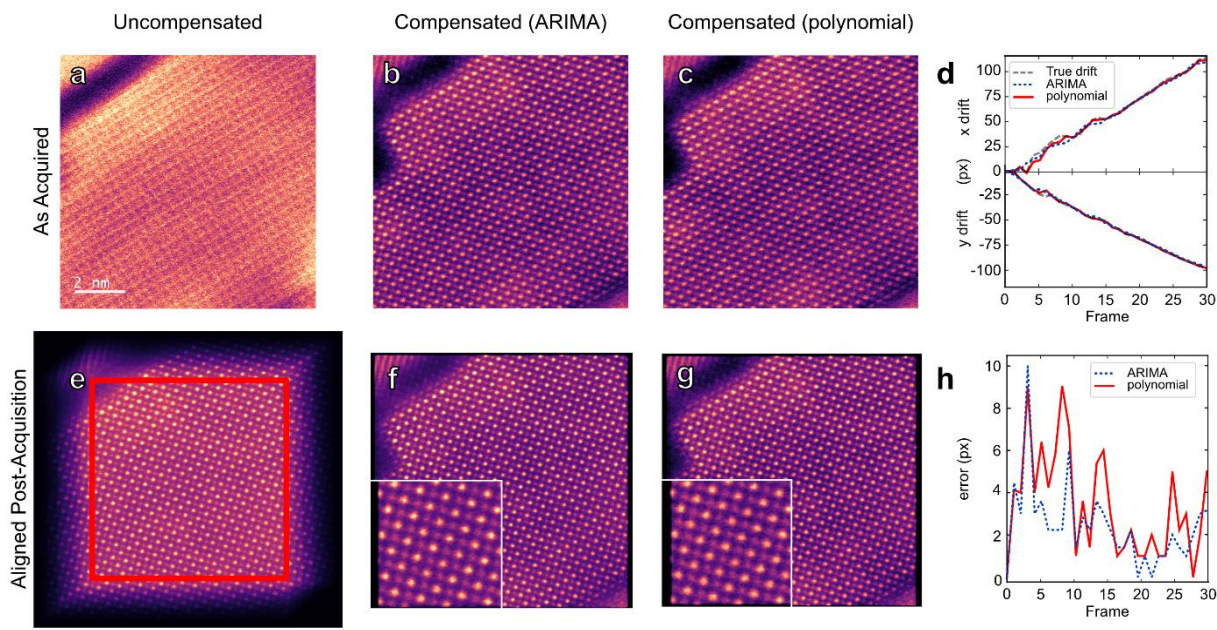

*Figure 3: Experimental 30 frame STEM acquisition, with three conditions applied synthetically, (a) no compensation, showing significant blurring from uncorrected physical drift, (b) live correction using ARIMA prediction model and (c) using a polynomial fit, with their average images shown. (d) shows compensated drift in x and y. (e-g) Frames above after post-processing alignment using SmartAlign. (h) Residual uncompensated drift, which was removed by post-processing alignment.*

### (2) Settling Time Implications

It is often necessary for the TEM operator to wait a significant time after sample insertion to let the sample. To quantify this experimentally, we measured the settling behaviour (drift-rate) decay for an uncorrected FEI Titan G2 operating at 300 keV observing a calibration grid. By capturing 160 frames with an interval of 30 seconds, the drift over 80 minutes was measured, with the results shown in Fig. 4. This profile allows us to determine how long a user would have to wait from sample insertion until certain imaging criteria have been met (for this microscope/holder combination).

For a single frame acquisition, the main way the image (or spectroscopy data) is degraded is by affine distortion, specifically shear. From experience, and the general quality of images in the literature, we set a limit of 3-5% displacement from expected position of a pixel for a frame to be deemed acceptable. Considering a hypothetical scan of a 10 nm FoV, with a 512×512 image with and pixel dwell time of 20 µs, the operator would require an acquisition time of 5.24 s. These acquisition parameters would be quite typical for a atomic resolution acquisition with high SNR. At this overall frame-time, flyback time is negligible and can be been ignored. Now, examining again our experimentally measured stage settling profile, for this scenario drift must be left to settle for between 40-60 minutes after sample insertion to achieve the acceptable 3-5% displacement.

By collecting multiple faster frames, affine distortion can be reduced but there will be a smaller common area remaining after alignment and cropping. For another hypothetical acquisition strategy, with the same FoV and total acquisition time, but now divided into 10 frames of 2 µs dwell time instead, 90% to 75% of the original area will be common to all frames after waiting 15-30 minutes after sample insertion. This roughly three-fold reduction in waiting time is a significant advantage of fast multi-frame acquisition over classical imaging, but some loss of common FoV cannot be avoided.

With live drift compensation, the limiting factor is maximum achievable deflection of the scan coils. Under the same conditions as above and imaging between 30-50% of the width of the largest possible scan, a delay of only 2.5-10 minutes is required after sample insertion. However, this settling time is unlikely to be the limiting factor, because aligning and focusing to atomic resolution often take significantly longer. Therefore use of predictive drift correction effectively eliminates the need for sample settling before high fidelity imaging can be performed.

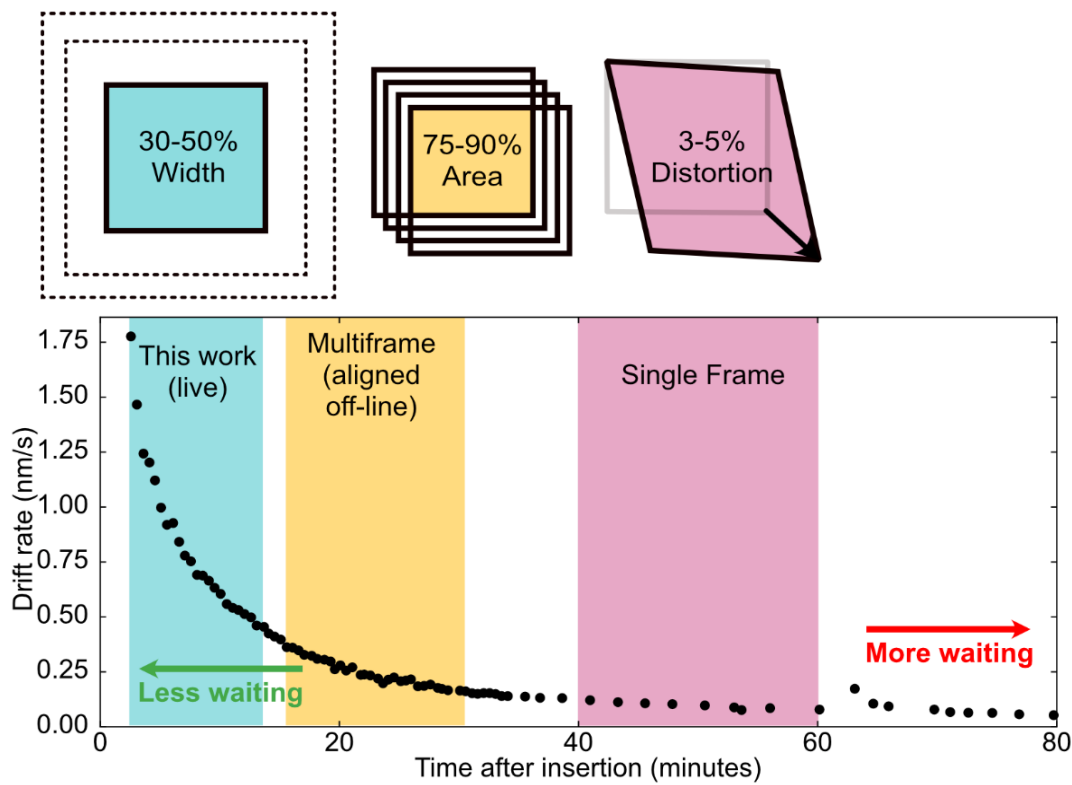

*Figure 4: Stage drift profile with time after holder insertion. Use of multi-frame acquisitions and live drift correction reduces user waiting time. Note, the uptick in drift at ~1 hour was due to refilling the liquid nitrogen in the anti-contaminator device.*

    *(3) Rigid Compensation*

The previous two examples show a synthetic acquisition and hypothetical settling times. To experimentally validate our live drift compensation acquisition for low-dose imaging at atomic resolution, an uncorrected FEI Titan G2 operating at 300 keV was used to image ⟨111⟩ yttrium aluminium garnet (YAG). All YAG samples were prepared by focused ion beam (FIB) lift-out using a Thermo Scientific Helios 5 Hydra CX Plasma FIB (PFIB)-SEM. Trenching and lift-out were performed using Xe, followed by final thinning with Ar.

First, two uncorrected frames with 1μs dwell time were captured as reference frames to measure drift, a single frame of which is shown in supplementary Fig. S2. Then 40 frames were captured, which alternated framewise between compensated and uncompensated scans. This acquisition approach is rather unusual but was done here to generate a genuinely fair comparison dataset, since the final outcome is a matched pair of series-acquisitions, one compensated and one not, with the acquisition order being interleaved in time such that the stage's drift-rate is effectively identical across the two series. Drift prediction was applied only to the compensated frame-set, determined using a polynomial fit over a rolling window considering the last 10 images into the past. 50% of the width of the maximum scan area of the instrument was used.

In the uncompensated acquisition when images are simply summed (Fig. 5a) there is a large amount of blurring to the point where no structure can be seen. To determine the drift present between frames, we used SmartAlign [5] to measure residual drift in the two image stacks. Without drift compensation, the sample moved 2.5 nm over the 20 frames. Consistent with this, Fig. 5d shows the common area after alignment is reduced to 71% of a full frame. This means a 29% of the time (and dose) was wasted on pixels that did not contribute to the final image, and the aligned image may have to be cropped further if a square frame is required for analysis or publication. If instead the aim is a full (512x512) frame after cropping, with this drift an initial frame 63% larger in area would have been necessary. This is compared to a crop losing less than 2% of the area for our drift compensated method.

Then by finding the shifts between the live drift compensated frames, it was found an average of 3.6 pixels or 0.66 Å of displacement was under- or over-compensated relative to the first frame. As opposed to the uncompensated case, the average of the drift compensated frames in Fig. 5b retains the atomic crystallography, though some small residual blurring is observed. For the highest resolution, Fig. 5e shows the live-compensated data finally offline-corrected with SmartAlign, with minimal loss of image area. As every other frame was an uncompensated frame and was not used for drift tracking algorithm, this residual displacement may have been improved still further if it were a genuinely drift compensated acquisition (without the extra interleaved frame times).

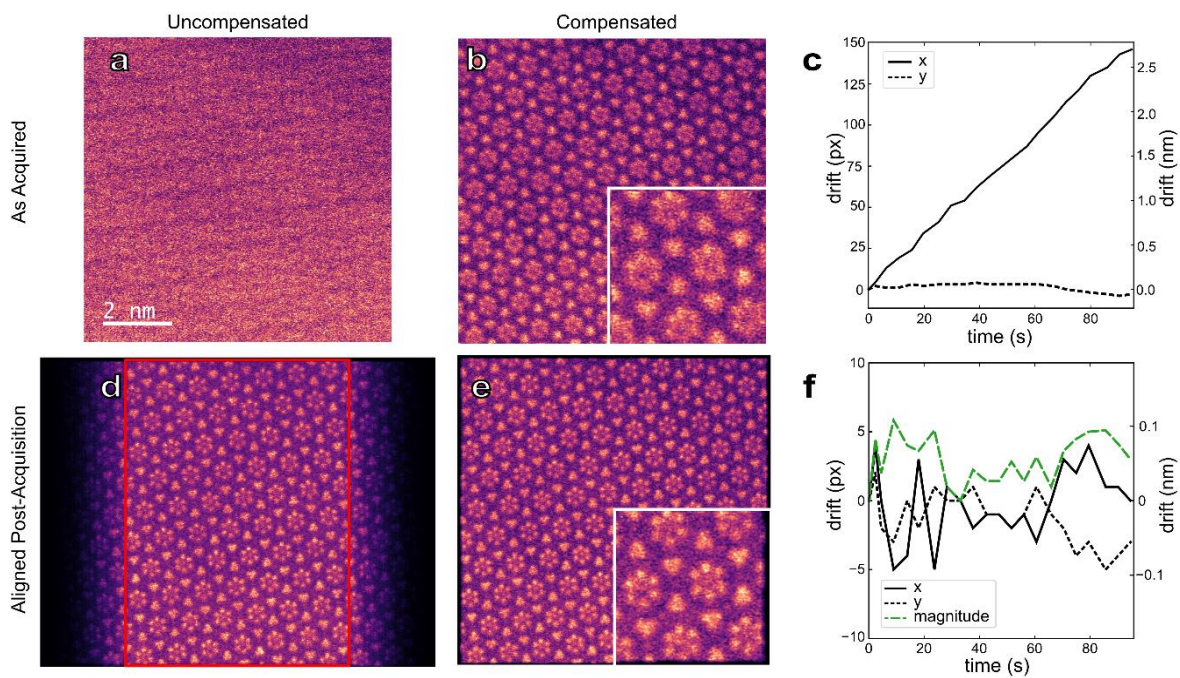

*Figure 5: Low-dose high-resolution 40 frame acquisition of YAG (111). (a) Sum of 20 frames that were not drift compensated and (b) sum of 20 frames that were drift compensated live during acquisition with inset showing twice magnification. (c) Drift occurring during the 40 frames. (d-e) Frames above after post-processing alignment, red outline showing common area. (f) Residual shift between frames, where the magnitude of residual drift (shown in green) is an average of 3.6 pixels.*

*(4) Affine/non-rigid Compensation*

Until now, a single (x, y) compensation has been applied to the scan coordinates of the whole frame. While this will reduce the translation between frames, there may still be remaining distortions within each frame. A changing drift velocity will shear images different amounts and thus produce a blur effect when stacked or aligned. A large and consistent drift will not blur the aligned stack, but will produce a shear in the image and make the interpretation of the crystallography incorrect. Since the drift function in Eq. 1 and 2 is just a function of time, and because a fully arbitrary scan generator is used, a pixelwise compensation can be applied.

To verify the ability to correct non-rigid distortions, 22 frames were captured of ⟨110⟩ YAG sample using a double aberration corrected JEOL ARM200F. After the initial 2 calibration frames, the acquired frames alternated between 10 frames with a rigid scan area offset, interleaved with 10 non-rigid compensated frames. In the latter, the polynomial function was used to predict the drift velocity to correct the position of *each pixel*, rather than only the whole frame. The polynomial was cubic after the $6^{th}$ frame and we used an exponential decay in weighting the fit into the past with a half-life of 3 frames.

The average of each of these sets of frames were then compared to a non-rigidly aligned rotating STEM acquisition [5], which is assumed to be the ground truth of interatomic distances and angles. The average of the rigidly compensated frames in Fig. 6a show a blurring towards the lower part of the frame, as the slowly changing *y*-drift seen in Fig. 6c causes different shear amounts in each frame, blurring the image. When the images are aligned in Fig. 6d the shear when compared to the ground-truth reference (highlighted in blue) is apparent, which arose from the large and consistent *x*-drift.

It was found that pixelwise non-rigid compensation in Fig. 6b & e reduced the blur from the changing *y*-drift and the shear from the large *x*-drift, as compared to the rigidly compensated set of frames. The shear was measured by use of a Radon transformation [34], which finds the angular position of lattice planes, and Fig 6f shows the difference in the angular position from the reference, and the rigidly compensated frames show a clear sinusoidal variance, which characteristic of shear [4], [35], with an amplitude of over 1%. Meanwhile, the pixelwise compensated frames have significantly reduced difference in the angular position.

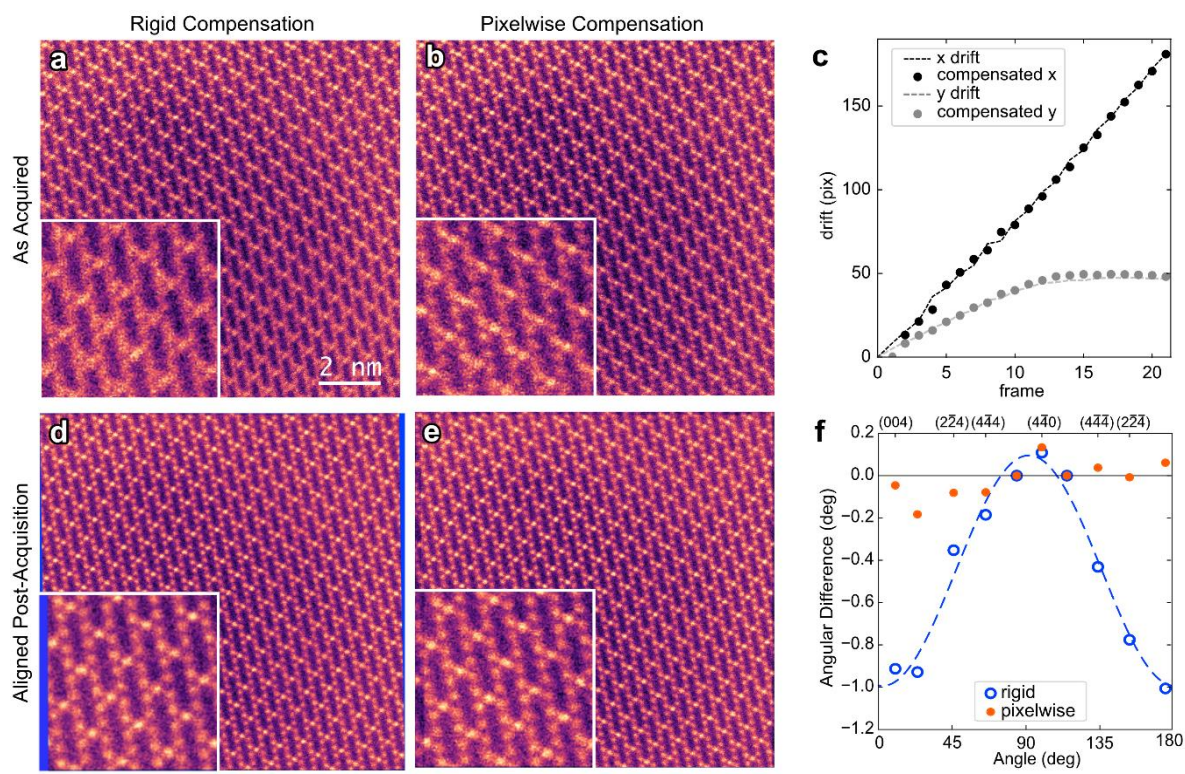

*Figure 6: 20 live drift compensated frames, alternating between rigid framewise compensation non-rigid/pixelwise compensation. (a) 10 compensated frames with rigid compensation applied and (b) 10 with pixel-wise compensation. (c) Drift trajectory during acquisition with large linear x drift and slowly changing y. (d&e) Frames above after post-processing alignment, blue background added to aid visualisation of the shear. (f) Measure of shear of images using Radon transform relative to rotating STEM.*

### (5) Rotating STEM non-rigid compensation

While predicting and compensating pixel positions reduces the shear in the image, it is difficult to confirm that it has been reduced to zero without comparison to ground truth. A common way of identifying and correcting this shear is to take successive frames with each scan direction rotating by 90° between each scan [4], [5]. This was done on a β-$Si_3N_4$ [0001] (SiN) on an uncorrected FEI Titan G2 operating at 300 keV. The SiN sample was prepared by FIB lift-out using a Thermo Scientific Scios Ga FIB-SEM.

Two separate rotating STEM acquisitions of 10 frames each were performed, one rigid and one non-rigidly compensated, and both acquisitions have comparable drift as seen in Fig. 7c. After registration (Fig. 7d), the rigidly compensated acquisition shows a blur around the outside, which is due to a periodically varying shear. Fig. 7b shows the non-rigidly compensated acquisition is uniform across the width of the frame before alignment, showing that the shear has been compensated, and there is no blurring effect around the border apparent in Fig. 7e. This means nearly distortion-free images were produced by combining rotating STEM and non-rigid drift prediction compensation.

In both acquisitions, there is a slight displacement that varies periodically every four frames as seen in Fig. 7f. This is because there is a temporal lag and therefore spatial shift in the scan, which will vary periodically in sign and direction with each 90° rotation [32]. Where drift is small or consistent, the periodic displacement can be added to the prediction function and compensated automatically, and an example of this is seen in supplementary Fig. S3. Once found, this offset can be used in subsequent acquisitions, though will be magnification- and scan speed-dependent.

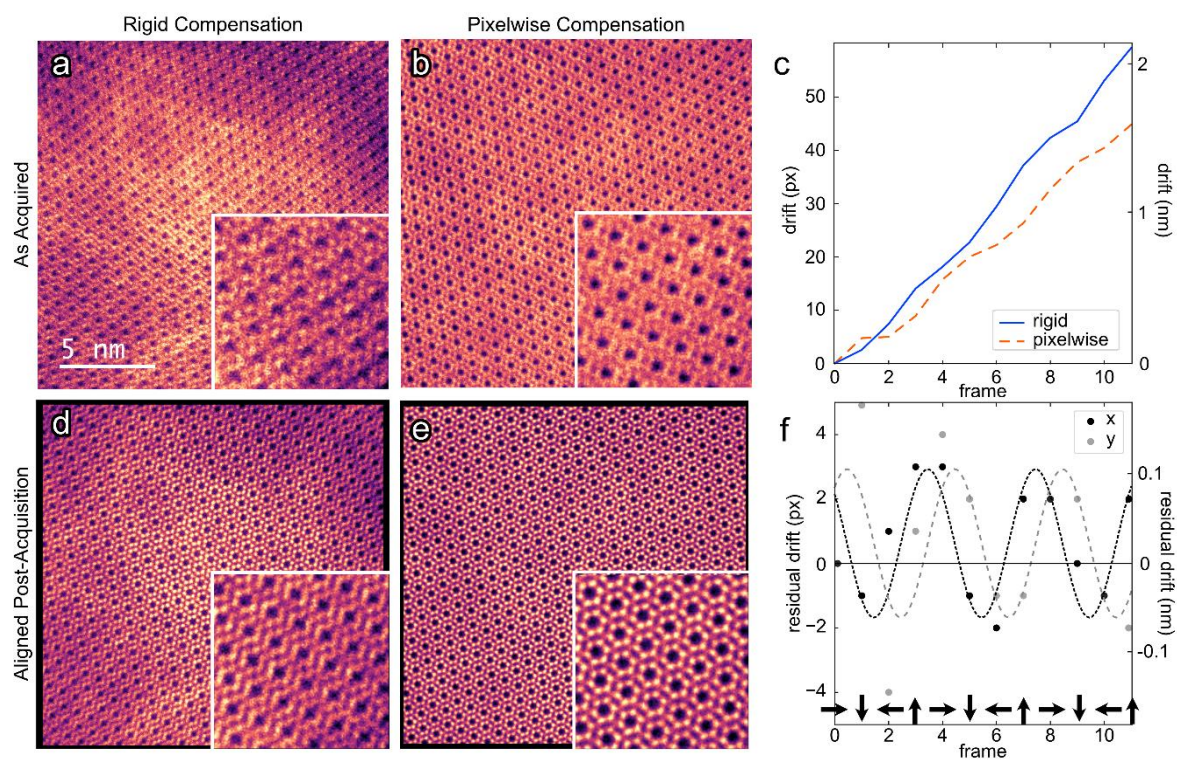

*Figure 7: Two separate compensated acquisitions with 10 frames each and a 90° rotation each frame. (a) Rotating acquisition with framewise/rigidly drift compensated acquisition. (b) Rotating with non-rigid drift compensated acquisition. (e) Comparison of drift in the separate rigid and non-rigid compensated acquisitions. (d-e) Frames above after post-processing alignment. (f) Residual displacement of Fig. 7b shows a periodic scan direction displacement (arrows showing fast scan direction).*

*(6) In situ Melting Experiment*

Heating experiments often produce significant amounts of drift due to thermal expansion in both the sample and its holder. This expansion typically makes imaging of a specific fixed area on the sample challenging. In addition, if there is a phase transition, particle ripening, mass-loss or any other significant changes to sample morphology, it may be difficult to accurately register images in post processing. Here, predictive drift compensation was tested in just such a challenging scenario.

Pure gold was sputter coated onto a DENS heating chip, with larger Au-Pd nanoparticles already present. The initial state is shown in Fig. 8a, after which the sample was heated from 100°C to 1000°C at a rate of 1 °Cs$^{-1}$. At 800°C it was held at this temperature for 1 minute (the maximum rated temperature for the heating chip) before continuing at the same ramp rate to 1000°C where it was held for a further 10 minutes.

Various significant sample changes were seen, with phenomena such as dewetting, melting, and Ostwald ripening occurring during the acquisition. Some example frames are shown in Fig. 8b-c (a further selection are shown in supplementary Fig S5). Across the whole 25 minute, 556 frame acquisition, drift was tracked and compensated with scan-offsetting, despite the large 451 nm physical drift and with drastic changes in sample morphology. The large structural change across the whole image series may mean that a cross correlation between first, middle and last frames would fail. However, the high frame rate meant that adjacent frames were structurally similar and cross correlation was accurate. Since a relatively short half-life of past frame weighting was chosen (3 frames), drift could be followed and compensated even with the large structural change observed from start to end of the experiment. Using this modest number of look-back frames in the analysis also reduced the number of cross correlations needed, ensuring the method could operate in real-time. This is a key consideration in in-situ imaging, since often reactions are not reversible and cannot wait for inter-frame calculations to be complete.

Drift (in x and y) was largest between 700-1000°C, where a large z-drift was also observed, and focus had to be adjusted manually to compensated for this. Drift was successfully tracked and compensated despite the changes in focus resulting from imperfect manual focus compensation. This is a further, albeit unintentional, test of the robustness of the approach to variations in image sharpness. Beyond 800°C the temperature of the heating chip is being pushed beyond the manufacturer's rated maximum, and was later found to have cracked (as shown in supplementary Fig. S6). Predictive drift compensation was able to successfully reduce drift in the acquisition to from 313 pixels (or 451 nm) to less than half a pixel on average (or 0.7 nm), and was 0 for most of the acquisition as seen in supplementary Fig. S7, except for during the sudden jump when the chip ruptured, but soon returned to the original FoV within a few frames.

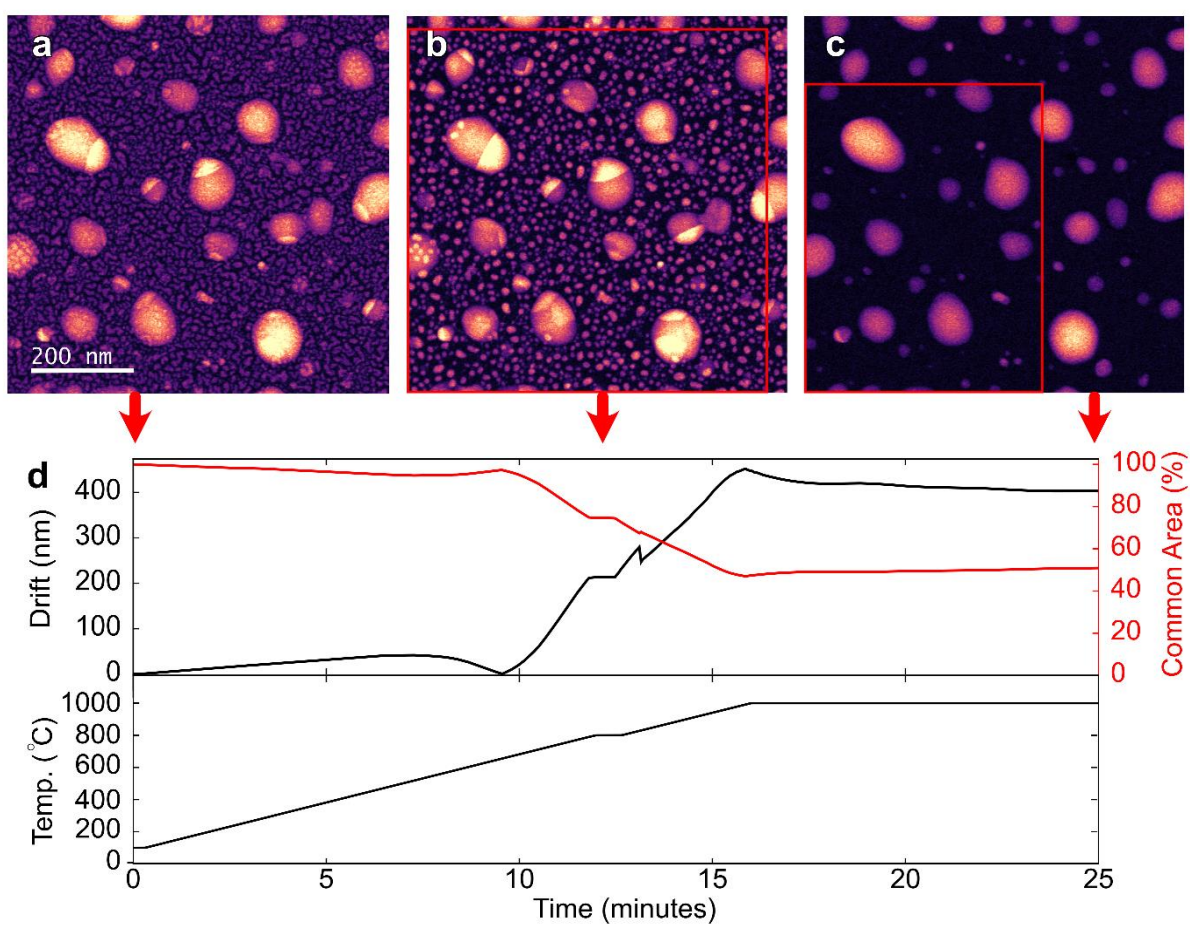

*Figure 8: (a) First, (b) middle and (c) last frame of in situ melting of Au. Red box corresponds to common area if drift was not compensated. (d) Graph of drift and resulting reduction in common area were drift not compensated in red and (below) temperature.*

# CONCLUSIONS

Stage or sample drift can make acquisitions inefficient and at worst corrupt whole series of data. Coarse corrections can sometimes be performed by hand, but this involves often-inaccurate stage adjustments, and precludes advances in instrument automation. Conventional post-acquisition drift correction can restore image quality but not the lost region of interest. Here we have demonstrated a *predictive* drift compensation method which identifies, extrapolates and compensates drift by adjusting the target scan coordinates pre-emptively for the next frame. This approach is applicable to any arbitrary scan generator that affords the operator sufficient control.

This work restores all but the second frame to the original area, with errors of only a few pixels, reducing the time the user needs to wait for the sample drift to decay. We have shown this is possible at both atomic resolution and for in-situ heating experiments at lower magnification, which means the operator is guaranteed the same physical area from first to last frame, reducing both acquisition time and beam damage. The method was shown to work on relatively low-SNR images, and the improved dose efficiency enables fast multi-frame imaging of beam sensitive samples.

We extended this method to predict and compensate individual pixel positions, which corrected shear in conventional multi-frame acquisitions and rotating STEM. This then allows the user to view live images with accurate and reliable feature positions even with significant drift, rather only being available in post-processing. There is the potential to expand this technique to pre-emptively correct higher frequency scan distortions or to correct across long spectrum image acquisitions after just a handful of fast image frames.

The algorithm used is available as open-source Python code, making this method accessible and replicable. This also makes it possible to reproduce predictive drift correction on other microscopy techniques, such as atomic force microscopy, scanning probe microscopy, SEM or confocal microscopy.

### AVAILABILITY OF DATA AND MATERIALS
The code used is available on Github. The data is available on request from the authors.

### ACKNOWLEDGMENTS
The authors acknowledge the facilities at the Trinity College Dublin CRANN Advanced Microscopy Laboratory and Warwick University Electron Microscopy Research Technology Platform. MM, MC and LJ acknowledge funding from the SFI-EPSRC CDT-ACM (grants 18/EPSRC-CDT-3581 and EP/S023259/1), MC acknowledges funding from Royal Society Tata University Research Fellowship (URF\R1\201318), EPSRC NAME Programme Grant EP/V001914/1 & Royal Society Enhancement Award RF\ERE\210200EM1, LJ and JJPP acknowledges funding from the Royal Society and Research Ireland grant URF/R/241034. A.S. and E.M. acknowledges EPSRC grant EP/V028596/1. J.G acknowledges HVM Catapult for partial funding of the WMG PFIB. Clive Downing is acknowledged for his immense help performing experiments.

### CONFLICT OF INTEREST
The authors declare that they have no known competing financial interests or personal relationships that could have appeared to influence the work reported in this paper.

# SUPPLEMENTARY FIGURES

## SUPPLEMENTARY FIGURE S1

For each frame *i* relative to frame *j*, the physical sample drift $d_{ij}$, (Fig. S1c) is found by taking the cross correlation of the two images, $c_{ij}$, (Fig. S1b) and adding the previously applied compensation to the scan coordinate, $b_{ij}$, (Fig. S1a)

$$d_{ij} = b_{ij} + c_{ij}$$

which when done for all image combinations will give a 2-D array of the relative shifts between each frame. For a number of images *n*, the size of this matrix grows with $n^2$. A large number of cross correlations will become computationally intensive.

There a number of ways the calculation can become more efficient, as seen in Fig. S1e. Half of the correlations are duplicates, as $c_{ij} = c_{ji}$, and $c_{ji} = 0$ where $i = j$. As only one row of the matrix needs to be added with each image, the number of computations in each iteration only grows with *n*. Finally, a cutoff in how far back in time to compute correlations can be set. The combination of these means the computation time comes to a constant value.

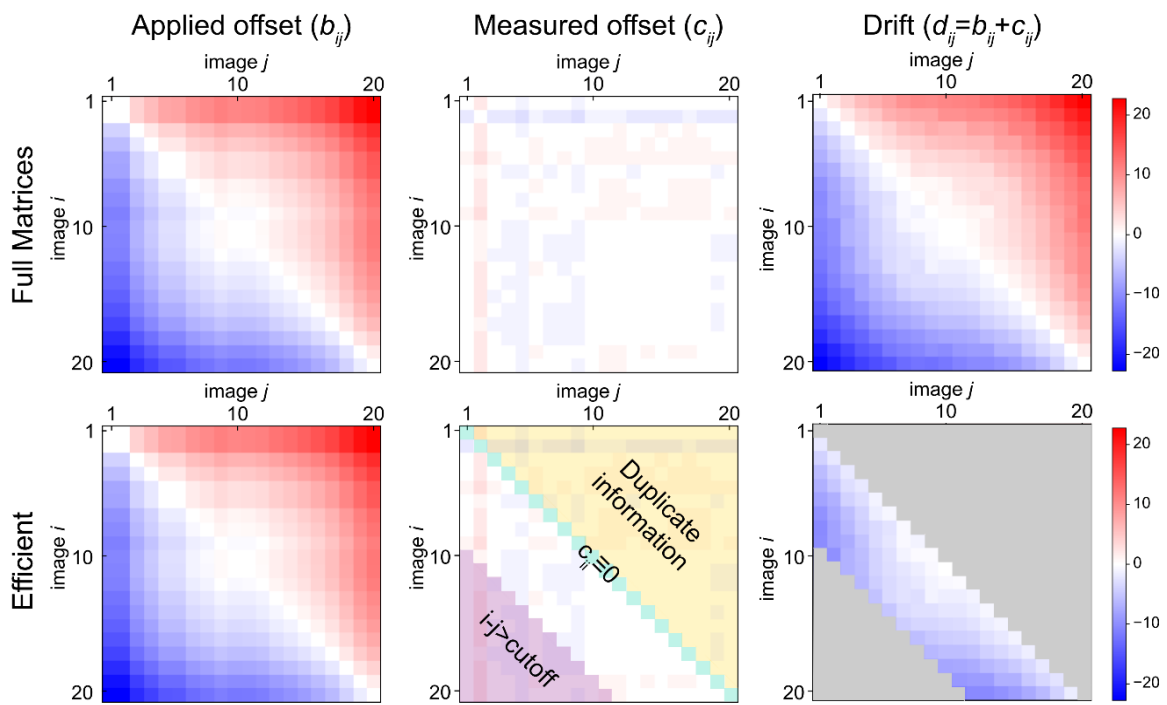

*Supplementary figure S1: Matrices of image i relative to image j of (from left to right) offset applied to scan coordinates, the measured displacement of images as found by cross correlation, and the estimate of true drift. Above are the full matrices with all redundant information. Below shows the how the number of correlations that computed can be reduced and the process sped up.*

# SUPPLEMENTARY FIGURE S2

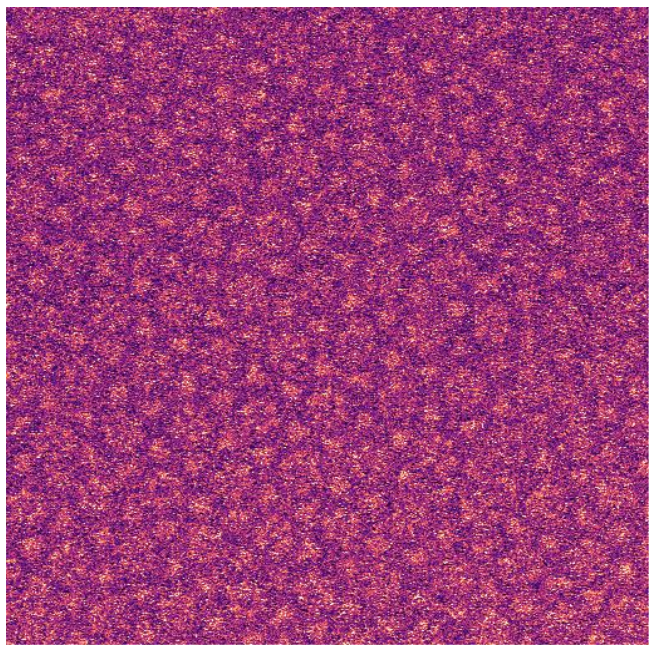

*Supplementary figure S3: The first frame of the drift compensated acquisition of ⟨111⟩ YAG, the ROI of which the other images are moved back.*

# SUPPLEMENTARY FIGURE S3

In order to emulate the process of navigating before acquisition then acquiring a multi-frame series, 10 fast frames were acquired but not compensated and the drift measured. A short pause was programmed in to emulate the time between navigation and acquisition. Then 12 frames were collected, where all frames were compensated to the first "slow" frame.

Note that no This was done under the same conditions and sample as figure 7, but without 90° rotation between each frame.

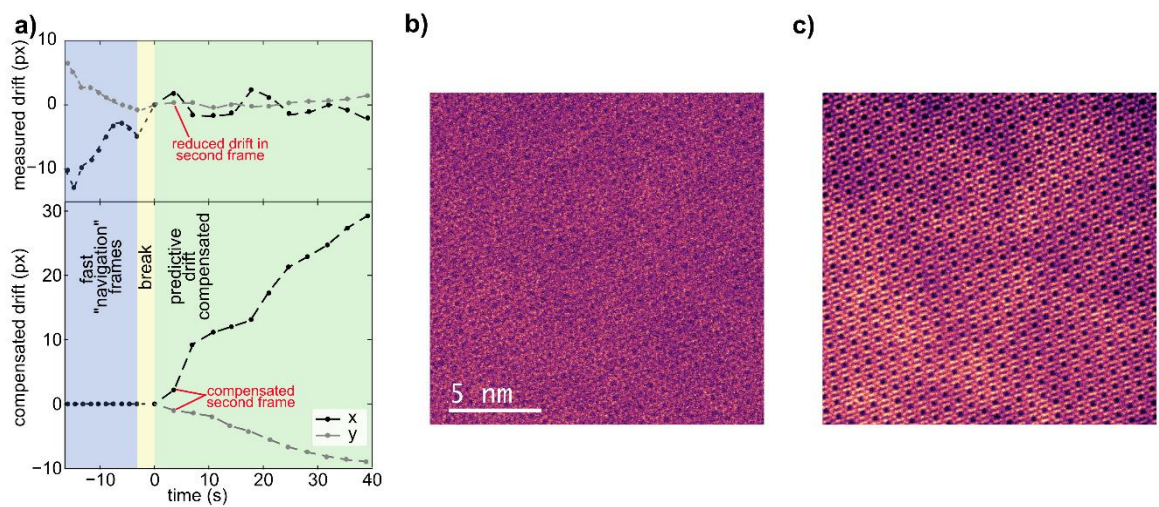

*Supplementary figure S2: (a) 10 fast frames are used to track drift, then a break, then 12 compensated frames were collected. Note the second frame was compensated using the fast frames and the first frame. (b) The first fast frame. (c) Average of 12 predictive drift compensated frames.*

# SUPPLEMENTARY FIGURE S4

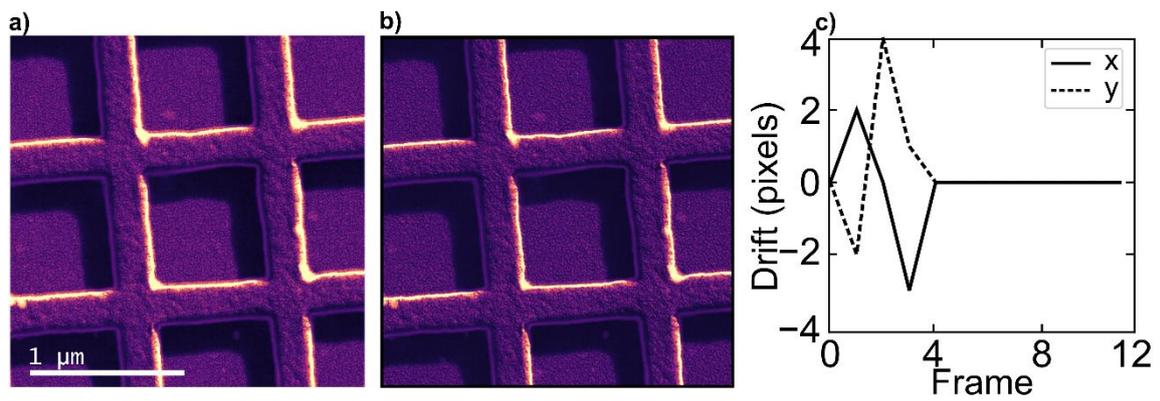

*Supplementary figure S3: 12 frames of low-magnification and therefore near-zero drift of a calibration grid with $90^0$ rotations between each frame. (a) Sum of frame showing slight "blurring" due to image shifts. (b) Aligned frames with small black border due to displacement. (c) Shift due to different scan directions can be automatically eliminated after the fourth frame.*

SUPPLEMENTARY FIGURE S5

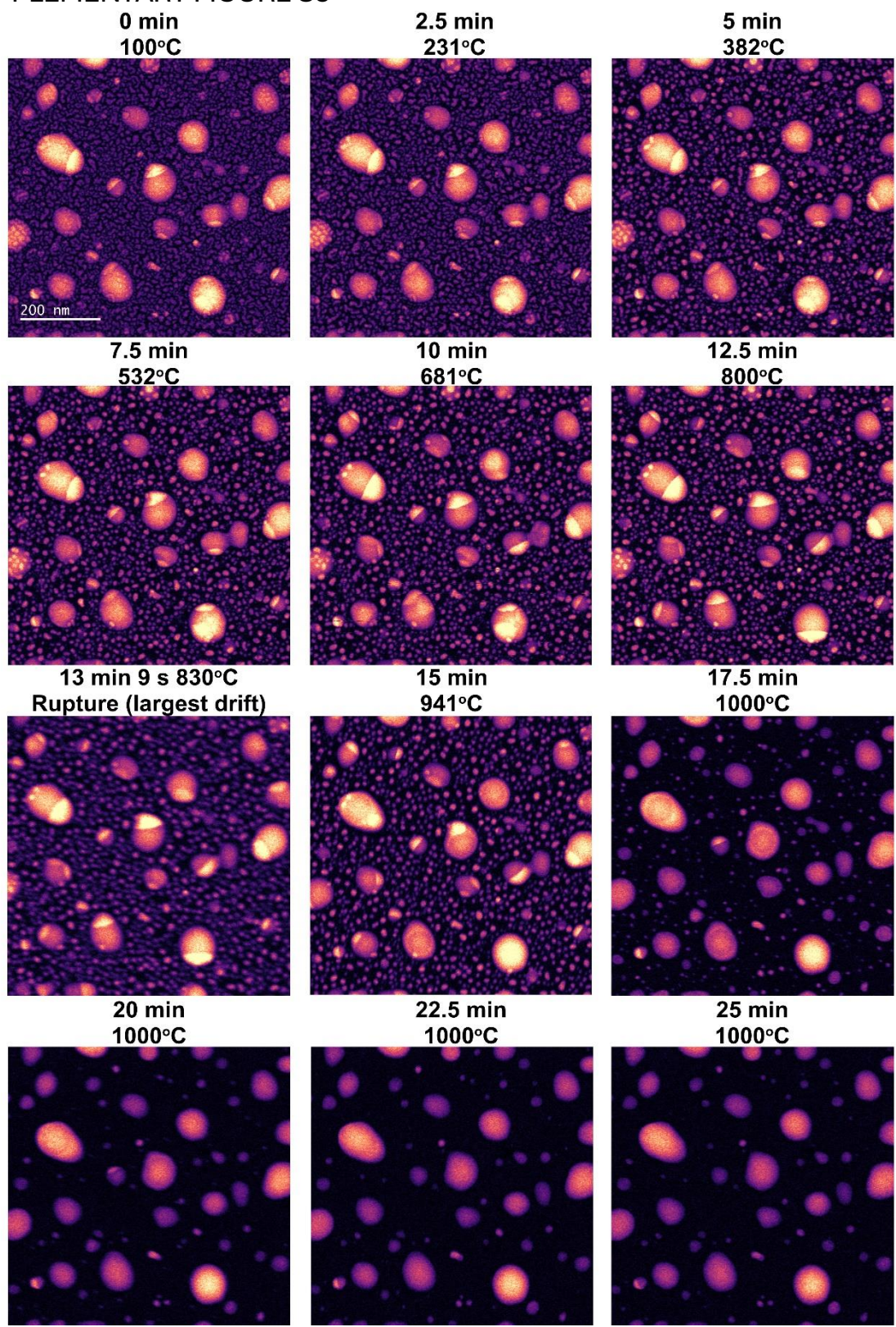

*Supplementary figure S5: 12 selected frames of melting sputtered gold, with frames spaced 2.5 minutes apart in time, except for the frame at which the heating chip ruptured at 13 mins and 9 seconds into the experiment.*

# SUPPLEMENTARY FIGURE S6

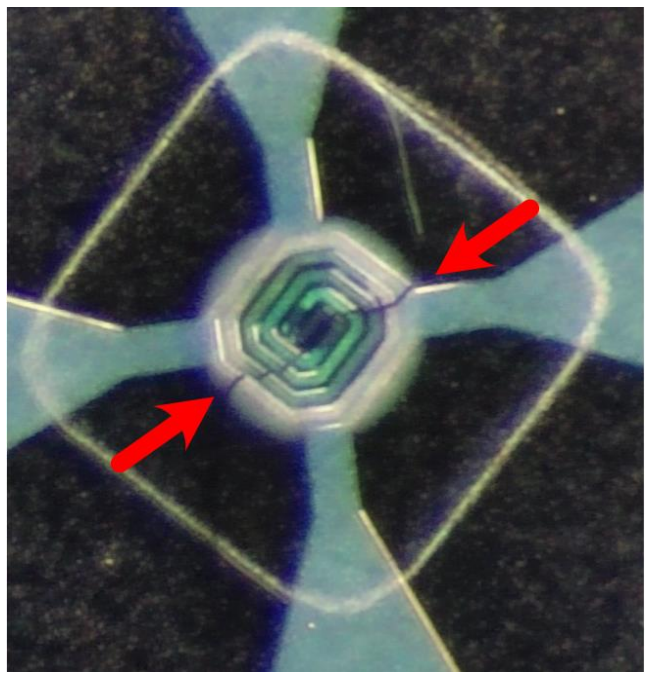

*Supplementary figure S6: Optical microscope image of DENS heating chip showing crack across device, which occurred during the heating experiment.*

# SUPPLEMENTARY FIGURE S7

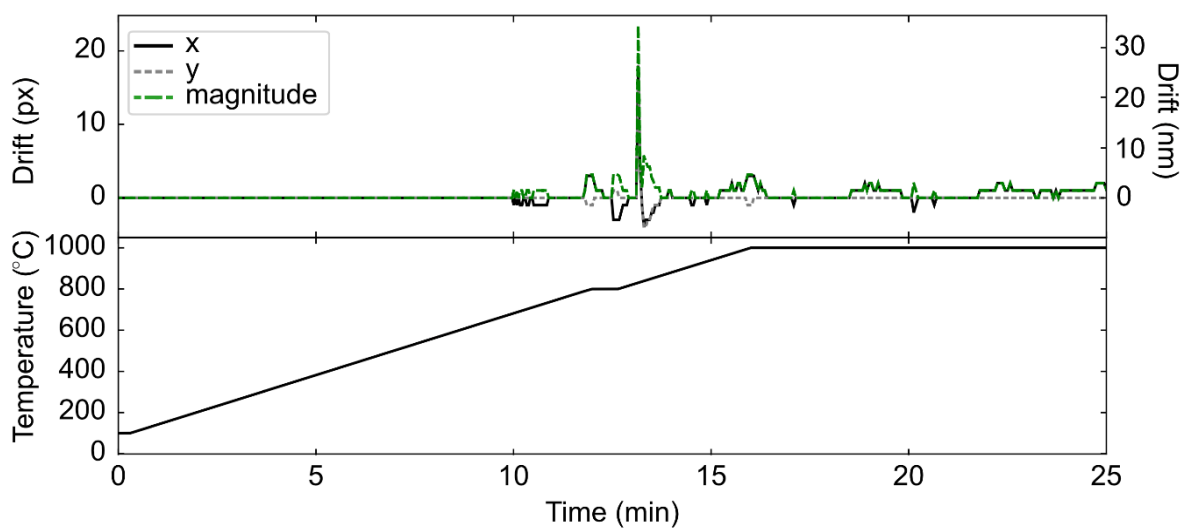

*Supplementary figure S7: (above) Residual drift as identified by SmartAlign [1] of the heating experiment and (below) corresponding temperature.*

References

[1] L. Jones, H. Yang, T. J. Pennycook, M. S. J. Marshall, S. Van Aert, N. D. Browning, M. R. Castell, and P. D. Nellist, Smart Align—a new tool for robust non-rigid registration of scanning microscope data, Adv Struct Chem Imag **1**, 8 (2015).